\documentstyle[twocolumn,epsfig]{mn} 
\newif\ifAMStwofonts

\def\cm3{\rm cm^{-3}}

\ifoldfss  
  \ifCUPmtlplainloaded \else 
    \NewTextAlphabet{textbfit} {cmbxti10} {} 
    \NewTextAlphabet{textbfss} {cmssbx10} {} 
    \NewMathAlphabet{mathbfit} {cmbxti10} {} 
    \NewMathAlphabet{mathbfss} {cmssbx10} {} 
  \fi  
  \ifAMStwofonts  
    \ifCUPmtlplainloaded \else  
      \NewSymbolFont{upmath} {eurm10} 
      \NewSymbolFont{AMSa} {msam10} 
      \NewMathSymbol{\upi}     {0}{upmath}{19}  
      \NewMathSymbol{\umu}     {0}{upmath}{16}  
      \NewMathSymbol{\upartial}{0}{upmath}{40}  
      \NewMathSymbol{\leqslant}{3}{AMSa}{36}  
      \NewMathSymbol{\geqslant}{3}{AMSa}{3E}

       \let\le=\leqslant 
      \let\geq=\geqslant \let\ge=\geqslant 
    \fi 
  \fi 
\fi 
 
\ifnfssone 
  \newmathalphabet{\mathit} 
  \addtoversion{normal}{\mathit}{cmr}{m}{it} 
  \addtoversion{bold}{\mathit}{cmr}{bx}{it} 
  \newmathalphabet{\mathbfit} 
  \addtoversion{normal}{\mathbfit}{cmr}{bx}{it} 
  \addtoversion{bold}{\mathbfit}{cmr}{bx}{it} 
  \newmathalphabet{\mathbfss} 
  \addtoversion{normal}{\mathbfss}{cmss}{bx}{n} 
  \addtoversion{bold}{\mathbfss}{cmss}{bx}{n} 
  \ifAMStwofonts 
    \ifCUPmtlplainloaded \else 
      \UseAMStwoboldmath 
      \makeatletter 
      \new@mathgroup\upmath@group 
      \define@mathgroup\mv@normal\upmath@group{eur}{m}{n} 
      \define@mathgroup\mv@bold\upmath@group{eur}{b}{n} 
      \edef\UPM{\hexnumber\upmath@group} 
      \new@mathgroup\amsa@group 
      \define@mathgroup\mv@normal\amsa@group{msa}{m}{n} 
      \define@mathgroup\mv@bold\amsa@group{msa}{m}{n} 
      \edef\AMSa{\hexnumber\amsa@group} 
      \makeatother 
      \mathchardef\upi="0\UPM19 
      \mathchardef\umu="0\UPM16 
      \mathchardef\upartial="0\UPM40 
      \mathchardef\leqslant="3\AMSa36 
      \mathchardef\geqslant="3\AMSa3E 

       \let\le=\leqslant 
      \let\geq=\geqslant \let\ge=\geqslant 
    \fi 
  \fi 
\fi 
 
\ifnfsstwo 
  \DeclareMathAlphabet{\mathbfit}{OT1}{cmr}{bx}{it} 
  \SetMathAlphabet\mathbfit{bold}{OT1}{cmr}{bx}{it} 
  \DeclareMathAlphabet{\mathbfss}{OT1}{cmss}{bx}{n} 
  \SetMathAlphabet\mathbfss{bold}{OT1}{cmss}{bx}{n} 
  \ifAMStwofonts 
    \ifCUPmtlplainloaded \else 
      \DeclareSymbolFont{UPM}{U}{eur}{m}{n} 
      \SetSymbolFont{UPM}{bold}{U}{eur}{b}{n} 
      \DeclareSymbolFont{AMSa}{U}{msa}{m}{n} 
      \DeclareMathSymbol{\upi}{0}{UPM}{"19} 
      \DeclareMathSymbol{\umu}{0}{UPM}{"16} 
      \DeclareMathSymbol{\upartial}{0}{UPM}{"40} 
      \DeclareMathSymbol{\leqslant}{3}{AMSa}{"36} 
      \DeclareMathSymbol{\geqslant}{3}{AMSa}{"3E} 

       \let\le=\leqslant 
      \let\geq=\geqslant \let\ge=\geqslant 
    \fi 
  \fi 
\fi 
 
\ifCUPmtlplainloaded \else 
  \ifAMStwofonts \else 
    \def\upi{\pi} 
    \def\umu{\mu} 
    \def\upartial{\partial} 
  \fi 
\fi 

\title{Three-dimensional simulations of the interstellar medium in  
dwarf galaxies - II. Galactic winds}
\author[A. Marcolini, F. Brighenti and A. D'Ercole] 
       {A. Marcolini$^1$, F. Brighenti$^1$ and A. D'Ercole$^2$ \\ 
       $^1$ Dipartimento di Astronomia, Universit\`a di Bologna,
       via Ranzani 1, 44127 Bologna, Italy \\ 
       $^2$Osservatorio Astronomico di Bologna,
       via Ranzani 1, 44127 Bologna, Italy}
  
\date{Accepted ..., Received ...; in original ...}   
  
\pagerange{\pageref{firstpage}--\pageref{lastpage}}

\pubyear{2004}

\begin{document}

\maketitle

\label{firstpage}

\begin{abstract}
We study the hydrodynamical evolution of galactic winds in disky dwarf
galaxies moving through an intergalactic medium.
In agreement with previous investigations,
we find that when the ram pressure stripping does not disrupt the
ISM, it usually has a negligible effect on the galactic wind dynamics.
Only when the IGM ram pressure is comparable to the central ISM 
thermal pressure the stripping and the superwind
influence each other increasing the
gas removal rate. In this case several parameters regulate the
ISM ejection process, as the original distribution of the ISM and
the geometry of the IGM-galaxy interaction.
When the ISM is not removed by the ram pressure or the wind, it loses
memory of the starburst episode and recovers almost its 
pre-burst distribution
in a timescale of 50-200 Myr. After this time another star formation episode
becomes, in principle, possible.
Evidently, galactic winds are
consistent with a recurrent bursts star formation history.

Contrary to the ISM content, the
amount of the metal-rich ejecta retained by the galaxy is
more sensitive to the ram pressure action. Part of the ejecta 
is first trapped in a low density, extraplanar gas produced by the
IGM-ISM interaction, and then pushed back onto the galactic disc.
The amount of trapped metals in a moving
galaxy may be up to three times larger than in a galaxy at rest.
This prediction may be tested comparing metallicity of dwarf galaxies
in nearby poor clusters or groups, such as Virgo or Fornax,
with the field counterpart.
The sensitivity of the metal entrapment efficiency on the geometry
of the interaction may explain part of the observed scatter in
the metallicity-luminosity relation for dwarf galaxies.
\end{abstract}

\begin{keywords}
galaxies: clusters: general -- galaxies: dwarfs -- galaxies:  
kinematics and dynamics -- galaxies: starburst -- hydrodynamics: numerical.  
\end{keywords}

\section{Introduction}

Dwarf galaxies are key players in theories of galaxy formation.  In
the standard cold dark matter picture they are the first forming objects,
and larger galaxies are successively built by merging of these
small systems (Blumenthal et al. 1984). Given their very low
metallicity and small size, these poorly evolved objects are excellent
laboratories to investigate the feedback of starbursts on the
interstellar medium (ISM) and to study their chemical evolution. In
models of dwarf galaxies formation the feedback from supernovae (SNe)
and the consequent gas and metals loss is a crucial process (Dekel \&
Silk 1986, Dekel \& Woo 2003).  The impact of starbursts in local
dwarf galaxies is well studied observationally (Martin 1998, 1999;
see Heckman 2003 for a recent
review), but important theoretical questions remain unanswered. 

A critical open problem is given by the dwarf galaxies chemical
evolution. Chemical evolution models of blue compact galaxies (BCGs)
(Matteucci \& Tosi, 1985; Pilyugin 1992; Marconi, Matteucci, \& Tosi
1994; Bradamante, Matteucci, \& D'Ercole 1998; Larsen, Sommer-Larsen
\& Pagel 2001) indicate that the gas fraction-metallicity relations of
BCGs is not compatible with the closed box scenario, suggesting that
(differential) galactic winds carry away a large fraction of the
metals produced by the young stars.  Numerical simulations
(e.g. MacLow \& Ferrara 1999, D'Ercole \& Brighenti 1999, Strickland
\& Stevens 2000, Recchi et al. 2001, 2002) and analytic models (De
Young \& Heckman 1994) indeed show that metals are easily ejected in the
intergalactic medium (IGM).  However, the details of the interaction
between the metal rich hot gas and the cold interstellar medium or the
IGM are yet not well understood.
For instance, the
mixing timescale for the metals produced in the starburst with the ISM
is poorly known. Chemical evolution models often assume instantaneous
mixing but HII regions nearby star formation sites do not appear to be
enriched by the young stars which illuminate them (Kobulnicky \&
Skillman 1996, 1997, 1998).  This suggests that most of the freshly
synthesized heavy elements do not mix right away with the surrounding
ISM and instead reside for timescales $> 10^7$ yr in the hot ($\ge
10^6$ K) phase, where they are indeed observed (Martin, Kobulnicky \&
Heckman 2002).  The question is then whether the hot gas leaves the galaxy
and enriches the IGM or whether it eventually cools and mixes with the ISM
on long timescales. The low but non-negligible metallicity of dwarf
galaxies suggests that some form of mixing is effective. Possible
mixing mechanisms are molecular diffusion (Roy \& Kunth 1995,
Tenorio-Tagle 1996, Oey 2003), condensation through thermal conduction
(McKee \& Begelman 1990) or turbulent mixing, likely the most
important one (Bateman \& Larson 1993, Roy \& Kunth 1995). The close
correlation between X-ray and $H\alpha$ emission also indicates some degree
of thermal mixing between hot and warm gas (e.g. Lehnert, Heckman \&
Weaver 1999, Strickland et al. 2002).

Babul \& Rees (1992) and Tenorio-Tagle (1996) proposed a
scenario in which the metal enriched superbubble powered by the wind
is confined by
a relatively high-pressure medium and then pushed back into the galaxy.
Babul \& Rees (1992) suggest that the confinment
is exerted by the IGM (ICM) of the group (cluster) to which the galaxy
belongs.  
This scenario was successively investigated in more
detail by Murakami \& Babul (1999), who estimate that the time-scale
for the superbubble to collapse is of the order of a few $10^7$
yr. In the model proposed by Tenorio-Tagle (1996, see also Silich \&
Tenorio-Tagle 2001), instead, the superbubble is
halted by a hypothesized gaseous halo surrounding the galaxy. 
The hot gas in the bubble then cools
and falls back onto the galaxy in a time-scale $\sim 1$ Gyr, much
longer than in the models by Murakami \& Babul
(1999). While in their simple form these scenarios
are probably too effective in retaining heavy elements (the low metal
content of dwarf galaxies implies significant metal loss), some
``weaker'' version, where only a small but non-negligible fraction of
the metal rich gas is able to cool, might help explaining the observed
metallicity of dwarf galaxies.

A further complication is given by of ram pressure effect if, as
expected, the galaxy moves through the IGM. It is
difficult to  predict the final effect of the interaction of a
galactic wind with  the IGM. The ram pressure  could be synergic to
the SN heating in removing the ISM from the  galaxy. Gas lifted above
the galactic plane by the SN energy may  be dragged away by the IGM
flow. Ram pressure effects were considered by Murakami \& Babul
(1999) in a number of 2D numerical simulations. These authors
find that when the ram pressure of the IGM is larger than its thermal
pressure, the expanding shell driven by the galactic wind is deformed,
fragmenting into dense clouds eventually dragged away from the
galaxy. Analogously, in the model by Tenorio-Tagle (1996) the ram
pressure may strip away the hypothesized extended, loosely bound ISM,
removing the medium confining the superbubble and the
metals contained in it.  On the other hand, may also happen that,
depending on the inclination angle between the galactic plane and the
orbital plane, at least one lobe of the expanding superbubble can be
squashed back on the galaxy by the ram pressure.  In this case the
enrichment process would be more efficient compared to the case where
ram pressure stripping is negligible.

Motivated by the above arguments, we have studied the interaction 
of starbursting dwarfs with the surrounding IGM, and how such 
interaction influences the evolution of the superwind powered by the 
stellar burst. In a previous paper (Marcolini, Brighenti \& D'Ercole 
2003, hereafter Paper I) we investigated the effect of the ram pressure alone
on the ISM of
disky dwarf galaxies.
Contrary to most of the other papers devoted to this argument, we
considered a ram pressure typical of galaxy groups rather than of rich
clusters, because most dwarf galaxies are found in the environment of loose
groups (Tully 1987). In the present paper we follow 
for a long timescale (500 Myr) the evolution
of a galactic wind originating at the galactic centre.
We focus in particular on the ejection of the ISM and of the 
metals synthesised in the starburst, and how their circulation is
influenced by the environment and by the ram pressure.

\section{the model}  
  
\subsection{Galaxy models and IGM parameters}  
 
We consider three galaxy models, described in detail in Paper I, 
differing for their masses and indicated with SM (small), MD (medium) 
and LG (large). The gravitational potential is due to a spherical 
quasi-isothermal dark matter halo and a stellar thin Kuzmin's disk. 
As in Paper I we defined two regions: the {\it galactic region} and 
the {\it central region}. The first is representative of the whole 
galaxy and is defined as a cylindrical volume within $z<|z_{\rm 
gal}|$ and $R<R_{\rm gal}$ (where $z$ and $R$ are the usual 
cylindrical coordinates). The second, smaller region samples the 
central region where most of the stars are located ($z<|z_{\rm 
centr}|$ and $R<R_{\rm centr}$); the masses of the gas content in these 
two regions are indicated as $M_{\rm gal}$ and $M_{\rm centr}$, 
respectively. The masses of the dark halo, stellar disk and initial 
ISM (before stripping) are summarized in Table 1.

The ram pressure effects are studied assuming two sets of 
density-velocity combination for the IGM: the first with ($\rho_{\rm 
IGM},v_{\rm IGM})=(2 \times 10^{-28} \rm{g} \, \rm{cm}^{-3}, 200 
\;\rm{km} \, \rm{s}^{-1}$) called LO (low), and the second with 
($\rho_{\rm IGM},v_{\rm IGM})=(2 \times 10^{-27} \rm{g} \, 
\rm{cm}^{-3}, 400 \;\rm{km} \, \rm{s}^{-1}$) called HI (high).
The IGM temperature is $T_{\rm IGM} = 10^6$ K.
 
In Paper I we run 18 simulations of rotating dwarf galaxies undergoing 
ram pressure stripping, varying the galactic mass, the ram pressure 
strength, and the inclination angle $\theta$ between the galactic 
plane and galactic velocity.  We identified a particular model with the 
notation XX-YY-ZZ, where XX individuates the galaxy size (SM, MD, LG); 
YY expresses the angle $\theta$, and takes the values YY=00 for
edge-on models, YY=45 for $\theta=45^\circ$, and YY=90 for face-on
models. Finally, ZZ represents the value of the ram pressure; ZZ=LO 
for the weak ram pressure, ZZ=HI for the high one. In the present
paper we do not consider the models SM-HI (for every inclination
angle), MD-45-HI and MD-90-HI because they quickly lose their ISM
during the stripping phase, and no galactic wind can occur (cf. Paper 
I).
 
The initial conditions for the gas in the present simulations are 
given by of the final outputs of Paper I, where the gas distribution has 
been modified by the effect of the ram pressure stripping lasting for 
1 Gyr.  
 
\begin{table*}  
\centering  
\begin{minipage}{130mm}  
\caption{Initial galaxy model parameters}  
\begin{tabular} {|c|c||c|c|c|c|c|c|c|c|}  
\hline 
Model&  $M_*$ & $M_{\rm {h,tot}}$ &               
     $M_{\rm{centr}}$ $^{(a)}$ &  
     $M_{\rm{gal}}$ $^{(b)}$  & $R_{\rm{centr}}$ &  $|z_{\rm{centr}}|$  
     & $R_{\rm gal}$  &  $|z_{\rm gal}|$  \\   
 &  $(10^8 \,M_{\odot})$ & $(10^9\,M_{\odot})$  
     & $(10^7 \,M_{\odot})$ 
     & $(10^8 \,M_{\odot})$ & (kpc) & (kpc) & (kpc) & (kpc) \\  
\hline  
SM   & 0.6  & 0.76 & 0.9 & 0.5 & 1.0 & 0.5 & 4.0 & 1.0 \\  
MD   & 6.0  & 7.4 & 6.6 & 4.1 & 2.0 & 1.0 & 8.0 & 2.0 \\  
LG   &60.0  & 77.2 & 42.2& 33.1 & 4.0 & 2.0 & 16.0 & 4.0 \\  
\hline  
\end{tabular}  
\par\noindent 
$^{(a)}$ initial ISM mass content of the central region defined as a 
cylinder with $R<R_{\rm centr}$ and $z<|z_{\rm centr}|$.\\ 
$^{(b)}$ initial ISM mass content in the galactic region defined as a  
cylinder with $R<R_{\rm gal}$ and $z<|z_{\rm gal}|$.\\  
\end{minipage}
\end{table*}

\subsection{The starburst}

We assume an instantaneous burst of star formation which injects 
energy into the ISM for a period of 30 Myr, approximately the lifetime 
of an $8 M_{\odot}$ star, the smallest star producing a Type II SN. 
We assume that the starburst produces a steady mechanical energy input 
rate of $L_{\rm inp}=3.8 \times 10^{40} {\rm erg} \, {\rm s}^{-1}$, 
which is a mechanical power similar to the lower limit estimated for 
NGC 1569 (Heckman et al.  1995). The mass injection rate is assumed to 
be $\dot M = 3 \times 10^{-2} M_{\odot} \, {\rm yr}^{-1}$. The total 
energy deposited after 30 Myr is then $\sim 3.6 \times 10^{55}$ erg 
and the total mass returned to the ISM by stellar winds and SNII is 
$M_{\rm ej,tot}=9 \times 10^5 M_{\odot}$ (cf. D'Ercole \& Brighenti 
1999). According to Leitherer \& Heckman (1995) the assumed injected
energy corresponds to an instantaneous starburst of 
$\sim 2 \times 10^6$ M$_\odot$.

\subsection{The numerical simulations}

We solve the usual hydrodynamic equations, with the addition of a 
mass source term and a thermal energy source term; the 
injected hot gas expands to form the starburst wind with the 
appropriate mechanical luminosity $L_{\rm inp}$. The constant mass and 
energy source terms for unit volume are given respectively by $\dot 
\rho= \dot{M} / \cal V$ and $\dot  
\epsilon = L_{\rm inp} / \cal V$, where  
$\cal V$ is the volume of the source region, chosen to be a sphere 
with a radius of 80 pc and located in the galactic centre. In order to 
follow the circulation of the metals produced in the starburst we 
solve an additional continuity equation for the stellar ejecta density 
$\rho_{\rm ej}$ (which is then advected passively). 

The simulations shown in Paper I did not take into account radiative  
losses. In fact, it can be shown that such losses are generally  
not important  
during the gas stripping. During the wind phase,  
instead, the ISM compressed by the wind has rather short cooling times  
and radiative losses must be included in the models. We use a fit  
for the cooling function given in Sutherland \& Dopita (1993), for 
solar abundances.

In order to achieve a better spatial resolution we remapped the final 
outputs of Paper I on a finer grid, with twice the number of zones in 
every direction. In the present paper 
the simulations are calculated on a grid with 
$212^3$ meshes ($212 \times 212 \times 108$ meshes for edge-on models 
with planar symmetry). The grid size has a geometrical 
progression in all directions with central zones $\Delta 
x=\Delta y= \Delta z = 20$ pc, with size ratio between 
adjacent zones of 1.0425. 

\begin{table}
\centering
\begin{minipage}{70mm}
\caption{Characteristic ISM values at the beginning of the starburst}
\begin{tabular} {|l|c|c|c|c|}  
\hline 
Model & $\theta$  
& $M_{\rm{centr}}$ & $M_{\rm{gal}}$ & 
$\overline{R}_{\rm{ISM}}$  
 
\\ & $(^{\circ})$ & $(10^7 \,M_{\odot})$  
& $(10^8 \,M_{\odot})$ & (kpc) \\ 
 
\hline 
       & 0  &  0.88 & 0.30 &  2 \\ 
SM-LO  & 45 &  0.91 & 0.25 &  2 \\  
       & 90 &  0.74 & 0.19 &  2 \\ 
\\ 
       & 0  & 0. & 0. & 0 \\ 
SM-HI  & 45 & 0. & 0. & 0 \\  
       & 90 & 0. & 0. & 0 \\ 
\\ 
       & 0  &  6.80 & 3.77 & 8 \\ 
MD-LO  & 45 &  6.31 & 3.24 & 6 \\ 
       & 90 &  6.17 & 3.28 & 6 \\ 
\\ 
       & 0  & 3.84 & 0.66 & 2  \\  
MD-HI  & 45 & 0.   & 0.   & 0  \\ 
       & 90 & 0.   & 0.   & 0  \\ 
MD-HIbis& 0 & 9.97 & 3.44 & 3  \\ 
\\ 
       & 0  &  43.0 & 33.0 & 22 \\ 
LG-LO  & 45 &  42.8 & 32.9 & 19 \\ 
       & 90 &  42.2 & 32.6 & 18 \\ 
\\ 
LG-HIbis& 0 & 62.5 & 26.5  & 6 \\ 
LG-HI   & 45 & 25.7 & 3.3  & 2 \\ 
        & 90 & 10.2 & 1.0  & 2 \\ 
\hline  
\end{tabular} 
\end{minipage}  
\end{table} 
 
\section{Results}

\begin{figure*}
\begin{center}
\psfig{figure=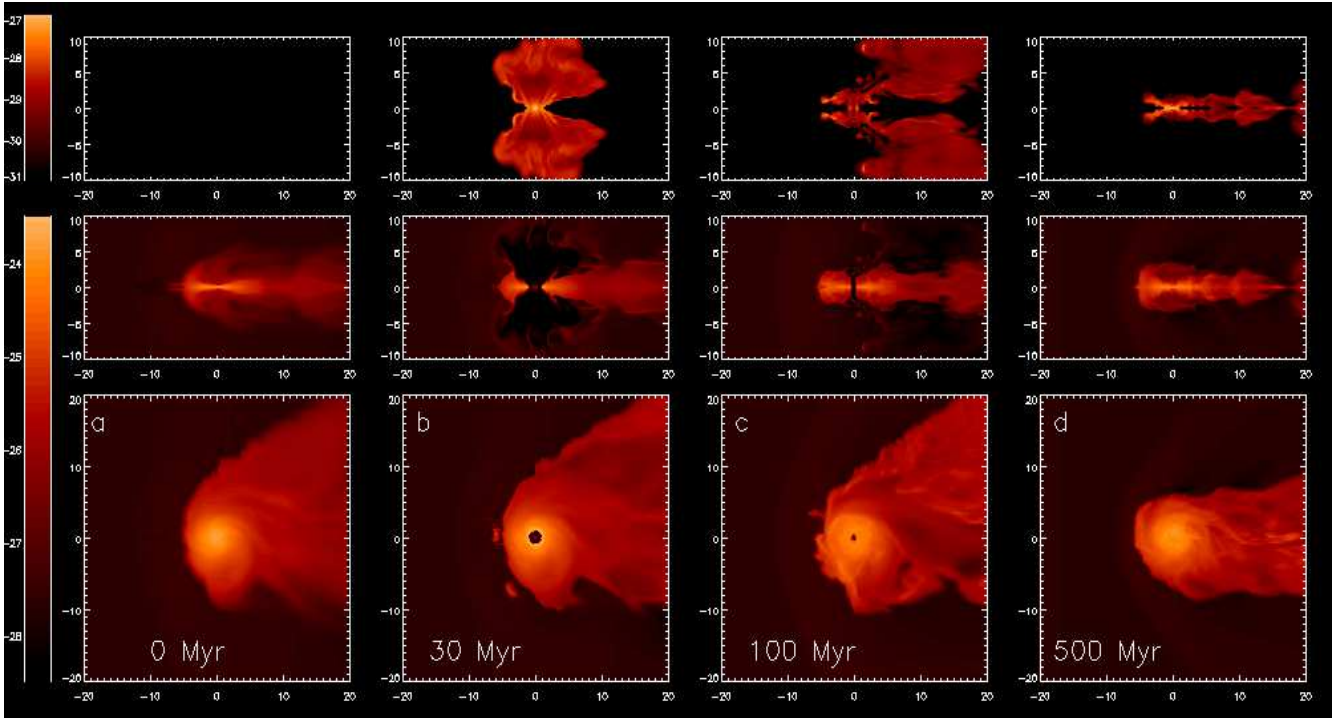}
\end{center}
\caption{Density distribution of the ISM and the ejecta in the reference  
model MD-00-LO at several times.  
The intermediate and lower rows illustrate the ISM distribution on the 
galactic plane ($z=0$) and meridional plane ($y=0$), respectively. 
The upper row refers to the ejecta distribution in the meridional plane. 
The distances are given in kpc.} 
\end{figure*}

Anticipating the results presented 
below, we stress the importance of the ISM flare at large radii
in our galaxy models. 
The flare of the initial gas distribution
is due to our assumption of isothermal
equilibrium of the unperturbed rotating ISM. Such a flare does not 
affect the evolution of the galactic winds occurring in the centre of 
a galaxy at rest. However, the ISM stripped from the leading edge of a 
galaxy moving through an external medium may form a 
sort of extraplanar gaseous halo surrounding 
the inner galactic region. Thus, the 
galactic wind propagation is not only affected by the direct impact of 
the ram pressure (at larger $z$), but also by the possible presence of 
such an halo (at smaller $z$). The halo is more substantial 
for larger galaxies and for larger values of ram pressure and, for 
geometrical reasons, this effect is more significant for edge-on 
models. 
 
Below we describe in some detail the gasdynamics of the models, paying 
a special attention to the galaxy MD, our 
reference model.

\subsection{The reference model}
 
\subsubsection{Model MD-00-LO: hydrodynamics}

In Fig. 1 several snapshots of the gas distribution 
for model MD-00-LO are shown
at different times; the second and the third rows 
refer to the ISM distribution in the meridional and galactic plane
respectively, 
while the first row refers to the SN-ejecta distribution 
in the meridional plane. The IGM enters the grid from the left boundary
and moves toward the right. 

The first column shows the gas density distribution just before the 
starburst episode ($t=0$).  
As the SNe start to explode,
a superbubble expands supersonically through the ISM, giving rise 
to the classical two shocks configuration. The reverse shock interacts 
with the freely expanding wind which is heated to $T \ge 10^7$ K.
As the outer shock propagates the ISM is compressed into a 
shell. This shell quickly breaks out and
accelerates along the $z$ direction 
as it moves through a less dense environment.
The bubble as a whole acquires the classical
bipolar shape. As a 
consequence of the acceleration, the shell becomes Rayleigh-Taylor 
(RT) unstable and fragments.  

The second column of Fig. 1 represents the wind evolution at $t=30$ 
Myr, the time at which the SNe activity ceases. At this time the central hole 
carved in the ISM reaches its maximum extension 
of $R \sim 1$ kpc; an hole of approximately this size has been 
effectively observed, for example, in Holmberg I (Ott et al. 2001, 
Vorobyov et al. 2003).  Along the $z$ direction the wind expands 
freely to much larger heights and the shape of the reverse shock is 
not spherical; similar structures are found by D'Ercole \& Brighenti 
(1999), Mac Low \& Ferrara (1999) and Strickland \& Stevens (2000)
for models with thin gaseous disks. The shell of shocked IGM, with
density $\rho \sim 6.3 \times 10^{-28}$ g cm$^{-3}$ and temperature
$T\sim  2.2 \times 10^6$ K, never cools to form a thin dense structure
as it  does when the superbubble is expanding within the ISM.
At $t=30$ Myr the outer shock  has reached a distance 
$z \sim 12$ kpc above the galactic
plane and  has a velocity of $\sim 230$ km s$^{-1}$ along the
$z-$axis. At this  stage the effect of the IGM ram pressure is not yet
substantial and the superbubble is still quite
axially symmetric  around the $z-$axis. As the energy injection stops,
the pressure of the hot gas decreases by expansion (and by
radiative losses occurring at the numerically broadened interfaces
between hot and cold gas) and the ram pressure drags
the gas expelled by the galaxy downstream.

\begin{table*}
\centering
\begin{minipage}{140mm}
\caption{Ejecta masses in the central region}
\begin{tabular} {|l|c|c|c|c|c|c|c|c|c|}
\hline
Model & \multicolumn{4}{c}{$M_{\rm ej, centr}^{\rm cold}$ $(10^3
M_{\odot})$} & & \multicolumn{4}{c}{$M_{\rm ej, centr}$ $(10^3
M_{\odot})$}
\\ & 10 Myr & 30 Myr & 100 Myr & 500 Myr & & 10 Myr & 30 Myr & 100 Myr
& 500 Myr \\
\hline 
LR,COOL     &  $29.9$  &  $38.3$
            &  $24.1$  &  $47.3$    &
            &  $55.1$  &  $65.1$
            &  $24.1$  &  $49.4$  \\
LR,NO-COOL  &  $6.88$  &  $5.86$
            &  $0.94$  &  $1.86$    &
            &  $35.0$  &  $35.8$
            &  $0.95$  &  $1.88$  \\
HR,COOL     &  $25.2$  &  $32.7$
            &  $25.2$  &  $42.9$    &
            &  $50.6$  &  $59.6$
            &  $25.3$  &  $43.0$  \\
\hline
\end{tabular}
\par\noindent
The total ejecta mass for $t\geq 30$ Myr is $9 \times 10^5$
M$_{\odot}$.\\
\end{minipage}
\end{table*} 

At $t=100$ Myr (Fig. 1, panels $c$) most of the gas at high $z$  is
moved behind the galaxy. The central hole in the ISM is
still present, but it will be filled up by the cold gas on the
galactic plane
which recollapses toward the centre. 
When this happens (at $t\sim 250$ Myr) another central starburst 
might occur (D'Ercole \& Brighenti 1999).

  The last column of Fig. 1 (panels $d$) shows the gas distribution at
$t=500$ Myr. The ISM in the central region has almost recovered its initial 
distribution, although the gas density at the centre is somewhat 
lower than that at $t=0$. This is due to the ejection of the low angular 
momentum-gas initially located in the very centre of the galaxy. The 
ISM which flows in the central region at later time has higher
angular  momentum and does not reach the very centre because of
angular  momentum conservation.    Overall, the galaxy has the classic
cometary shape, typical of  galaxies undergoing ram pressure stripping
(Paper I and references  therein). The tail becomes narrower with
time, an indication that  the system is not in a steady state.

Fig. 5 shows the evolution of $M_{\rm centr}$ and $M_{\rm gal}$
before the starburst (discussed in Paper I) and following it. The
temporary decrease of the ISM mass occurring at $\sim 1$ Gyr
(i.e. when the starburst occurs) in the central region is
due  to the galactic wind. As the energy injection stops, 
the mass gas increases
again. The gas mass inside the central region at the end of the
simulation is slightly larger than its initial value.  This is due to
gas ablated from the leading edge of the galaxy by the ram pressure
which later falls in the central region.

\subsubsection{Model MD-00-LO: SN ejecta evolution} 

In this section we investigate the fate of the metals injected by the 
SNe created by the starburst. This task, however, is severely 
hampered by the effects of the numerical diffusion, which artificially 
mixes the SNe ejecta and the cold ISM. This is a well known problem, 
discussed for example in D'Ercole \& Brighenti (1999), Recchi, 
Matteucci \& D'Ercole (2001) and de Avillez 
\& Mac Low (2002). 
Generally, the real mixing processes as 
molecular diffusion (e.g. Oey 2003), condensation by thermal 
conduction (e.g. McKee \& Begelman 1990) or turbulent mixing (Begelman 
\& Fabian 1990, Bateman \& Larson 1993) are less effective than the 
numerical diffusion which affects our simulations. Therefore, our 
calculations likely provide an upper limit to the amount of metals 
trapped in the ISM.
Thus, before to describe the evolution of the SN 
ejecta in model MD-00-LO, it is useful to make a brief digression on 
the influence of the numerical diffusion on the mixing of the hot 
metal rich material. 

To study quantitatively the numerical mixing we consider as a 
fiducial model a (2D) galactic wind 
in the reference galaxy described above, but now assumed 
at rest (i.e. no ram pressure) to simplify the interpretation 
of the results.
\begin{figure*}
\begin{center}
\psfig{figure=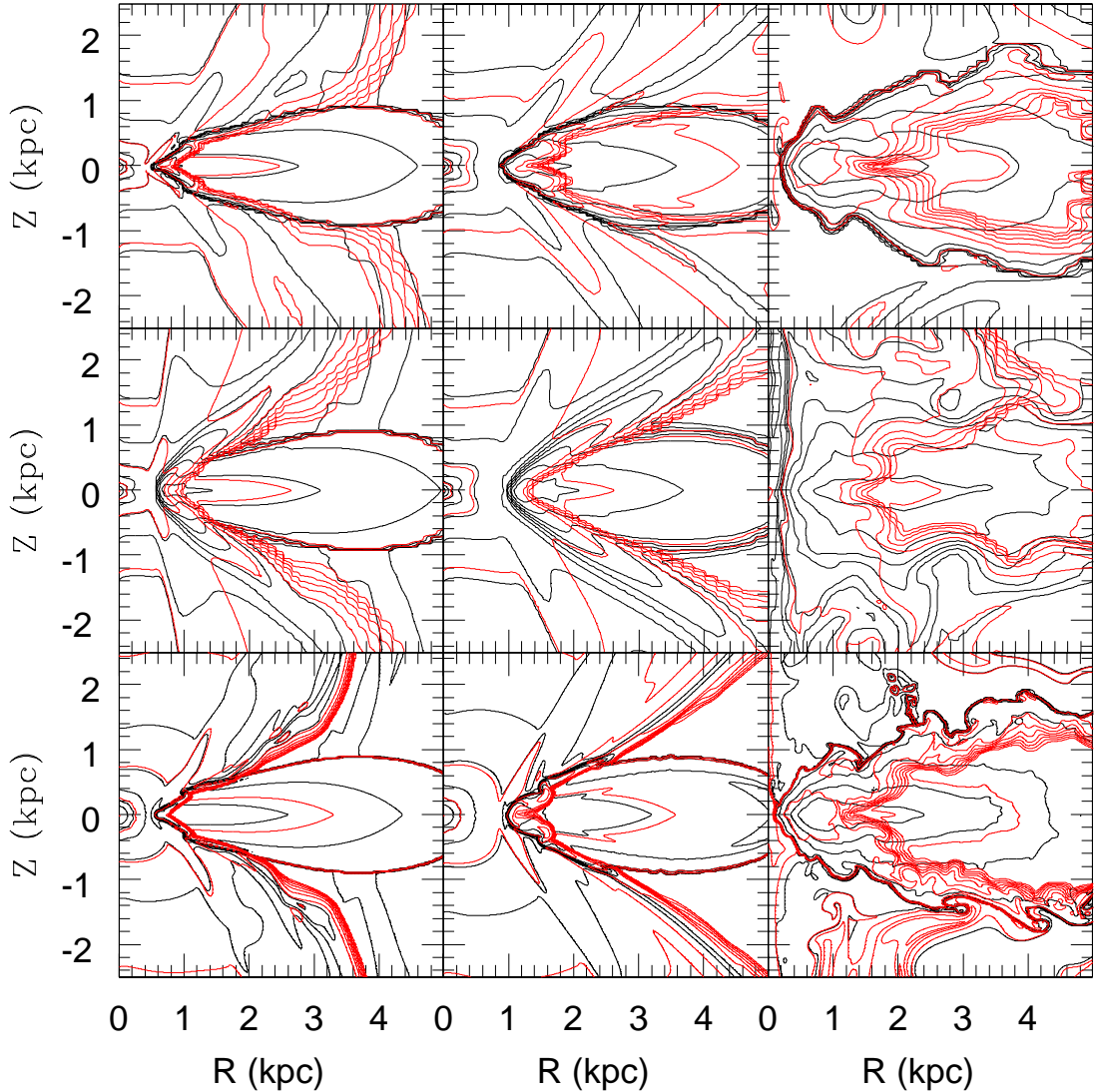}
\end{center}
\caption{Gas (black contours) and ejecta (red contours) distributions  
for the 2D model MD at rest for different grid resolutions, with and 
without radiative cooling. From top to bottom the rows refer to models
LR-COOL, LR-NOCOOL, HR-COOL respectively. From left to 
right the columns refer to $t=10$ Myr, $t=30$ Myr and $t=500$ Myr, 
respectively. Given the symmetry of the model each panel shows only 
half of the meridional plane. The $z$-axis coincides with the symmetry 
axis.}
\end{figure*}  
Fig. 2 illustrates the evolution in the meridional plane
of the gas density 
(black contours) and the ejecta density (red contours) for 
this model.  The top row shows the model calculated at the same 
resolution of models MD-00-LO, at $t=10$, 30 and 500 Myr.  This model 
includes radiative cooling and will be indicated in the following as 
LR,COOL. The three snapshots clearly show how the ejecta penetrates 
the cold ISM.  In Table 3 we list the cold ($T<10^5$ K) and the total mass 
of the ejecta ($ M_{\rm ej,centr}^{\rm cold}$ and $M_{\rm ej centr}$)
in the central region for the various models 
described in this section. In absence of numerical diffusion we would 
expect $M_{\rm ej, centr}^{\rm cold}= 0$, since the radiative cooling 
time of the ejecta is longer than the age of the wind. 

We find that after the energy injection stops (at $t=30$ Myr) 
essentially all the ejecta found in the central region is cold, as a 
result of numerical diffusion.  In the third row of Fig. 2 we show the same 
model calculated with higher resolution (model HR,COOL).  The linear 
grid size in the relevant region (say at $\sim 1$ kpc from the centre) 
is now about 3 times smaller than in model LR,COOL.  From Table 3 we
see that, somewhat unexpectedly, increasing the resolution does not
reduce the amount of the diffused ejecta.
We note that de Avillez \& Mac Low (2002) 
similarly found that the mixing timescale of chemical inhomogeneities 
in their models for the Galactic ISM was rather insensitive on the 
numerical resolution. 
  
The effect of the radiative cooling is instead crucial. In the second 
row it is shown the low resolution model calculated with the radiative 
cooling turned off (model LR,NO-COOL).
Now $M_{\rm 
ej,centr}^{\rm cold}$ and $M_{\rm ej,centr}$ are greatly reduced with 
respect to models LR,COOL and HR,COOL.  At late times, in models 
without radiative cooling the central region hosts less than 5 \% of 
the ejecta mass present in the analogous models which include 
radiative losses (a high resolution model without radiative losses,
not shown here,
gives similar results). Evidently when radiative losses are included,
the ejecta cools at the numerically 
broadened interfaces between hot and cold gas.
We must conclude that the majority 
of the ejecta trapped in the ISM in our models is due to the 
combination of numerical diffusion and radiative cooling. 
Thus, we consider the results of our simulations 
as an upper limit of the mixing efficiency.

\begin{figure*}
\begin{center}
\psfig{figure=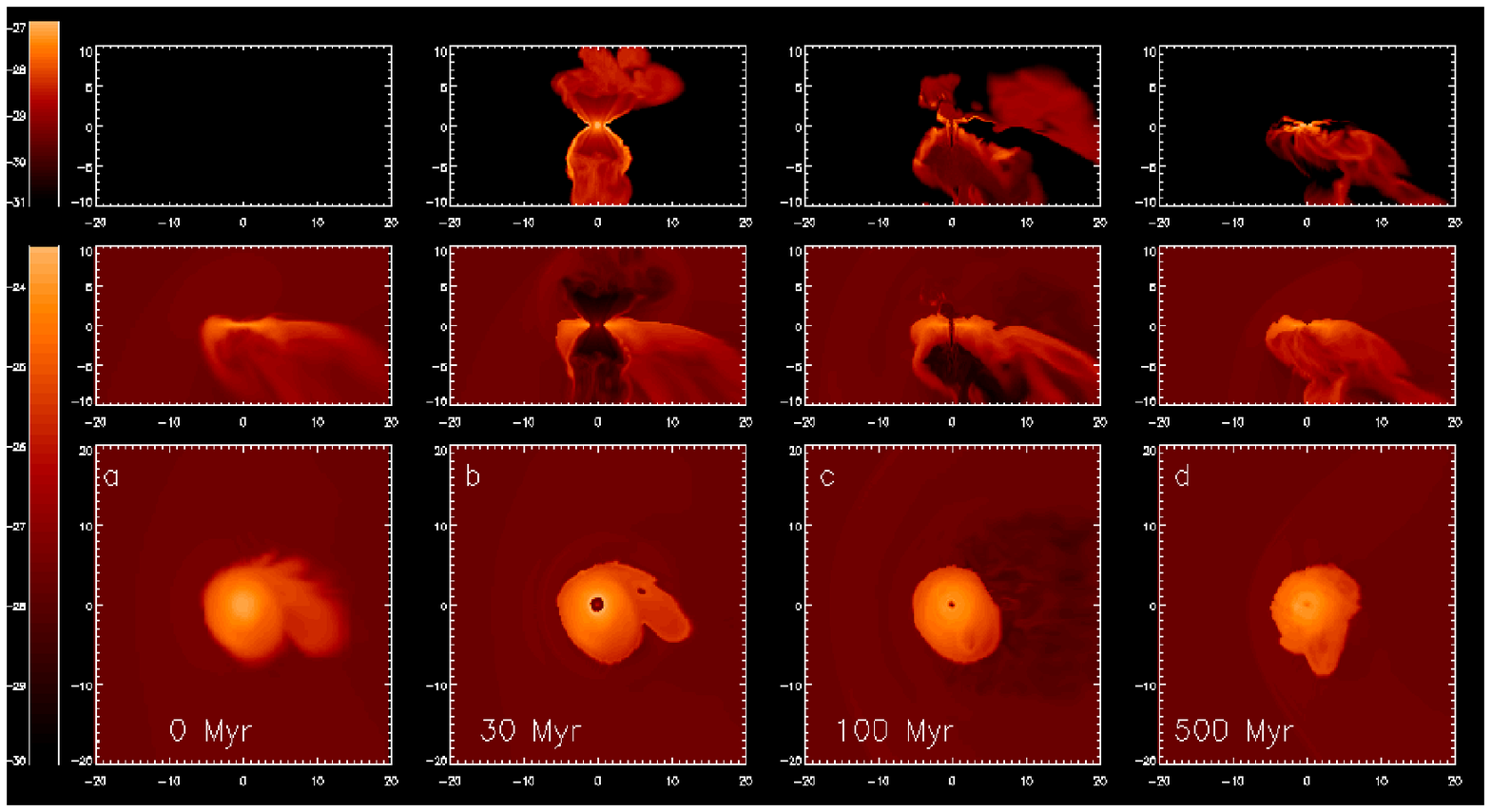}
\end{center}
\caption{The same as in Fig. 1 but for the model MD-45-LO.}
\end{figure*}

We now come back to model MD-00-LO. The upper panels of Fig. 1
show the evolution of the ejecta density in the plane $x-z$, while the 
time evolution of the ejecta masses $M_{\rm ej, centr}$ and $M_{\rm 
ej, gal}$, defined similarly to the gas masses in the section 2.1, is 
shown in Fig. 7 (solid line). Initially, the mass of the ejecta increases as 
it is delivered by SNe. 
Successively, at 
$t>30$ Myr, the ejecta close to the galaxy continues to move 
vertically leaving the central region and causing the minimum in 
$M_{\rm ej, centr}$ seen in Fig. 7 at $\sim 100$ Myr.  The ejecta 
which reaches large heights ($z\geq 2-3$ kpc) is effectively dragged 
away by the ram pressure (Fig. 1, panels $c$).  However, some of the
ejecta surrounding the galaxy is pushed back onto the galactic plane 
together with the ISM stripped from the leading edge of the 
galaxy, as discussed above. As a consequence, $M_{\rm ej,centr}$ increases
again before to attain a nearly constant value at $t\sim 200-300$ Myr. 
At the final time $t=500$ Myr about 4\% of the metals produced in the 
starburst is within the central region.

The evolution of $M_{\rm ej,gal}$ shows less variation with respect to 
$M_{\rm ej,centr}$ because of the larger volume involved (see Fig. 
7). After $t=30$ Myr about a constant fraction $\sim8$\% of the 
total ejecta mass is located in the galactic region. Almost all the 
ejecta inside both the galactic and the central regions
has cooled off and condensed onto the ISM for $t \geq 100$ Myr. It is located
along the edges of the ISM (Fig. 1) and it is unaffected by
the ram pressure. The vast majority of the ejecta is thus lost
definitively by the galaxy and pollutes the IGM.

\subsubsection{Models MD-45-LO and MD-90-LO}

The first column in Fig. 3 represents the gas distribution of
model MD-45-LO at the beginning of the burst.  The IGM enters the grid
from the left and from the top and moves in the diagonal direction. 
Downstream (i.e. for $z < 0$) the ablated ISM forms a gaseous
halo close to the galaxy. Thus, as the wind starts to blow, the upper
lobe of the superbubble expands through a less dense medium ($\rho= 3 \times
10^{-28}$ g cm$^{-3}$)
compared to the gas surrounding the lower lobe
($\rho= 1-2 \times 10^{-27}$ g cm$^{-3}$). As a consequence it
expands faster despite the decelerating action of the ram pressure.
During the SNe activity ($t \le 30$ Myr) the ram pressure is 
comparable to the superbubble pressure at large distance ($z > 4$ kpc),
where the symmetry between the two lobes breaks.
The density distribution at $t=30$ Myr is illustrated in
the second column of Fig. 3. From the ejecta distribution (upper
panel) one can see that the downstream lobe has reached
larger heights ($\sim -15$ kpc) compared to the upper lobe ($\sim 12$
kpc). The last two columns of Fig. 3 show the successive
evolution of the ISM and the ejecta. The latter is progressively
dragged downstream
and at late times no ejecta is present ``above'' the galaxy.  As for
model MD-00-LO, the ISM loses memory of the starburst episode and 
almost recovers its initial distribution at $t\sim 200 $ 
Myr.

\begin{figure*}
\begin{center}
\psfig{figure=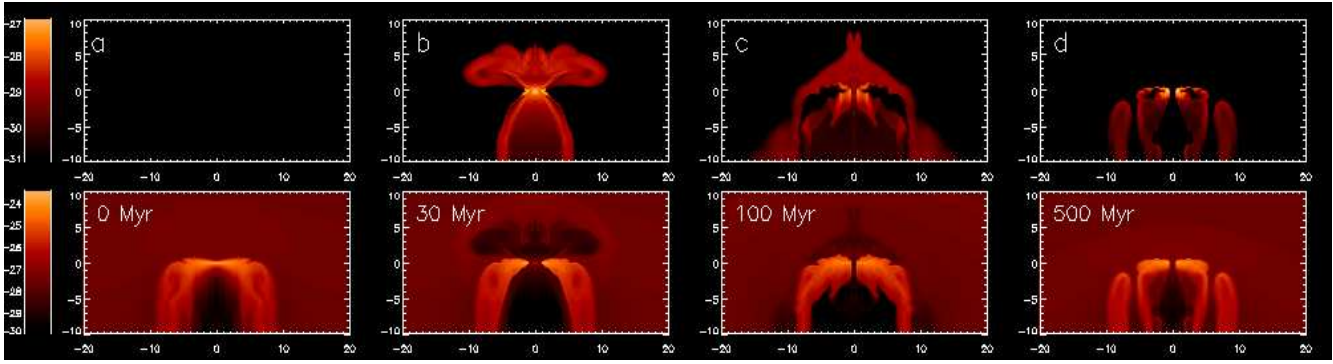}
\end{center}
\caption{Density distribution of the ejecta (upper panels) and the
ISM (lower panel) in model MD-90-LO in the meridional plane at several
times. The distances are given in kpc.}
\end{figure*}

Fig. 4 shows the gas evolution of the model MD-90-LO. The IGM 
enters the grid from the upper grid boundary. Given the symmetry of 
this configuration with respect to the $z-$axis, we have run this 
simulation on a 2D grid in cylindrical coordinates, with the same 
resolution as the previous calculations. In this model the superbubble 
lobe expanding upstream is smaller than the opposite lobe because of 
the ram pressure. At $t \sim 30$ Myr the outer shock propagating 
upstream overtakes the 
bow shock created by the motion of the galaxy (located at $z \sim 10$ 
kpc) and reaches the maximum distance
$z\sim 14$ kpc at $t\sim 40$ Myr.  Upstream, the 
density and temperature of the shocked IGM at $t=30$ Myr are $\rho 
\sim 6\times 10^{-28}$ g cm$^{-3}$ and $T \sim 2.5\times 10^6$ K, 
respectively.  In the downstream direction, the forward shock exits 
the boundary of the grid at $t\sim 20$ Myr.
The reverse shocks are located at $z\sim 3$ kpc upstream and at 
$z=-8$ kpc downstream and the shocked gas has temperature $T\sim 2-4 
\times 10^7$ K. On the galactic plane ($z=0$) the outer and reverse 
shocks have a radius of $\sim 1$ and $\sim 0.7$ kpc, respectively. The 
density and temperature of the shocked wind are $\sim 2.5 \times 
10^{-28}$ g cm$^{-3}$ and $7\times 10^7$ K,
respectively. 

As the SNe activity ceases, the ejecta expelled upstream 
is pushed back toward the galaxy. However, most of it moves around the 
galaxy and is lost downstream. The ISM collapses 
back in the central region at $t\sim 100$ Myr. However, a central 
small hole is always present and some of the external gas moving toward 
the galaxy goes through this hole which has a radius of $\sim 300$ pc 
(see also Quilis, Moore \& Bower 2000).

The evolution of $M_{\rm centr}$ and $M_{\rm gal}$ is quite similar
for all models MD-LO (Fig. 7), although $M_{\rm gal}$ assumes
slightly larger values for the edge-on model.
Also the behaviour of $M_{\rm ej,centr}$ and $M_{\rm ej,gal}$
is similar for all models MD-LO, with no clear dependence
on $\theta$. This point will be discussed further in
section 5.2.

Model MD-90-LO, being calculated on a 2D grid, is computationally 
inexpensive and it is therefore well suited for a numerical convergence 
study. We have recalculated 
this model on a finer grid with
central zone size of $10 
\times 10$ pc$^2$, and size ratio between adjacent zones 
1.00447. The total number of zones is thus $1200 \times 600$.
Although, as expected, many small scale 
structures are present in the high resolution simulation,
the overall dynamical evolution of the galactic wind is very 
similar to that of the low resolution model. 
The evolution of $M_{\rm centr}$, $M_{\rm 
gal}$ and the ejecta masses is also very similar at high and low 
resolution. At $t=500$ Myr, $M_{\rm ej, centr}$ and $M_{\rm ej, gal}$ 
for the low resolution model is only $\la 20$\% larger than in the 
high resolution one, consistently with the experiments on the 
numerical diffusion described in section 3.1.2. We conclude that the 
relatively low numerical resolution of our models does not affect the 
global evolution of the ISM and the SN ejecta.

\subsection{Models MD-HI}  
 
We now discuss the evolution of a galactic wind in the same galaxy 
model described in the previous section, but in the case of a higher 
ram pressure (HI). As shown in Fig. 5 and discussed in Paper I, 
for model MD-00-HI the ram pressure phase before the starburst results 
in a loss of $\sim 80$ \% of the original ISM in the galactic 
region. In the central region the effect is milder and only $\sim 40$ 
\% of the gas is removed.  Because of the lower ISM content, the 
galactic wind is now able to remove completely the remaining ISM after 
$t\sim 200$ Myr since the burst occurrence. Its worthwhile to note 
that, even in absence of a central starburst, the ISM would have been 
completely removed by the ram pressure at $t \sim 700$ Myr (i.e. after 
1.7 Gyr from the beginning of the stripping). Therefore, the wind 
anticipates the total stripping of the gas by $\sim 500$ Myr.  
A scenario in which both SNe energy and ram pressure stripping 
act together to remove the ISM is argued by Gallart et al. (2001)
for the Phoenix dwarf galaxy.

\begin{figure*}
\begin{center}
\psfig{figure=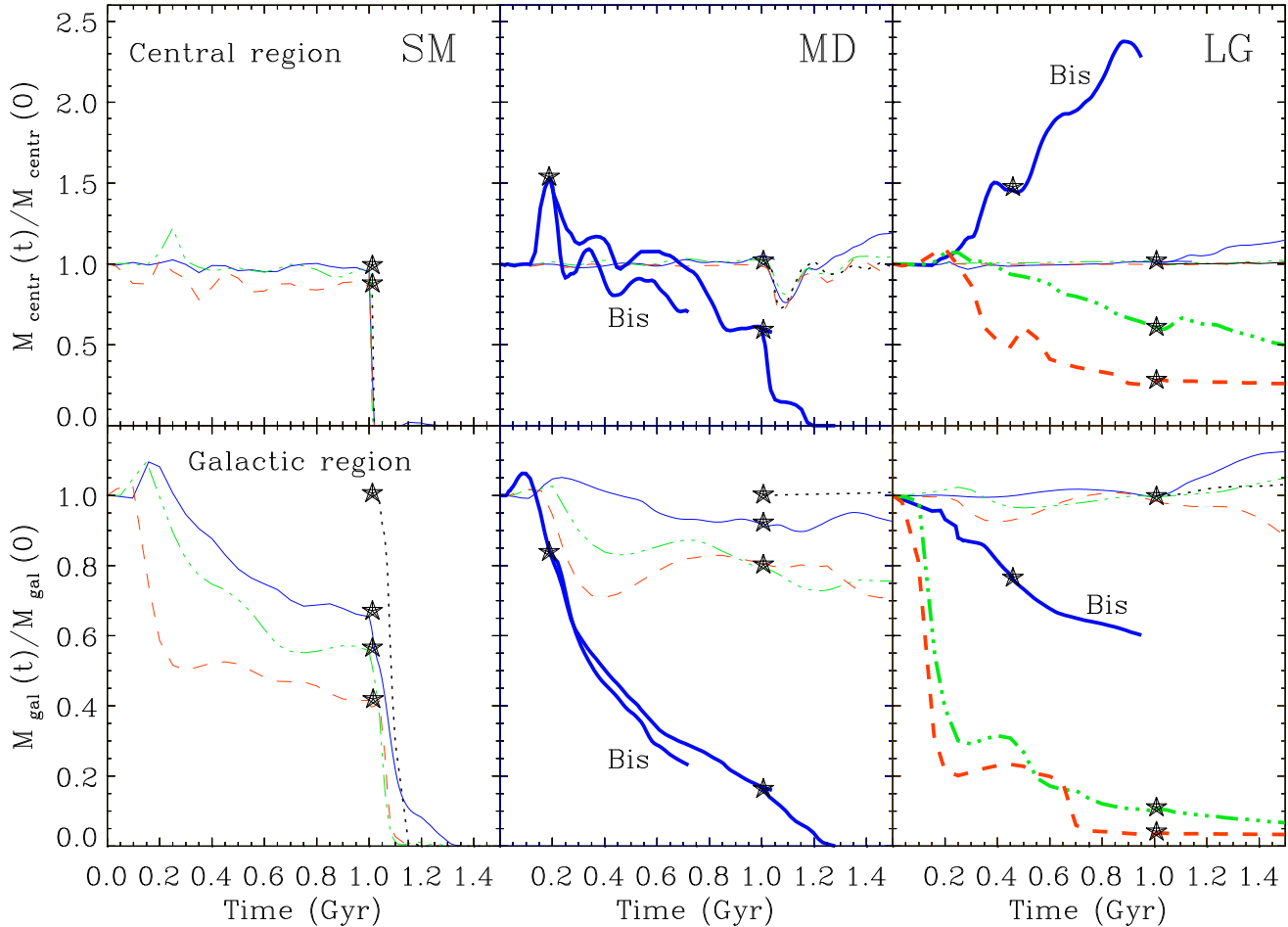}
\end{center}
\caption{Evolution of the cold ($T<3 \times 10^5$ K) ISM mass 
in the ``central region''  
(upper panels) and in the ``galactic region'' (lower panels).
From left 
to right the couples of lower and upper panels refer to SM, MD and LG 
models respectively. 
Solid blue lines: edge-on models; dashed-dotted green lines: $45^{\circ}$ 
models; dashed red lines: face-on models; dotted lines: REST models. 
Light and heavy lines refer to LO and HI ram pressure models, 
respectively. Being the FIELD models very similar to the REST models 
we do not report them in the picture for the sake of clarity. The 
evolution of the masses is reported over a time interval of 1.5 Gyr: 
during the first 1 Gyr (with the exception of models Bis, see text) the 
masses evolve only under the action of the ram pressure (Paper I). The 
occurrence of the starburst is indicated by a star.} 
\end{figure*} 

The duration of the ram pressure stripping phase, 1 Gyr, is
arbitrary.
Such a time interval is comparable 
or longer than the crossing time of a typical group (a few $10^8$ yr). 
Therefore, if the starburst 
episode is triggered by tidal interactions with other galaxies, or by 
the ram pressure itself (see Paper I), it may occur after a shorter 
lapse of time. In this case the superbubble can interact with a more 
massive ISM and its evolution can change somewhat. In order to 
investigate this point, we run a model MD-00-HIbis, identical to
MD-00-HI, but taking as initial condition the ISM
configuration at 190 Myr since the beginning of the stripping 
phase. At this time the ISM mass content in the central region is 
maximum, $M_{\rm centr} \sim 1.0 \times 10^8$ $M_{\odot}$ 
(see Fig. 5), and possible
differences in the bubble evolution are expected to be larger. Fig. 5
shows the evolution of $M_{\rm centr}$ and $M_{\rm gal}$ for both 
model MD-00-HI and model MD-00-HIbis.  In model MD-00-HIbis the more 
massive ISM prevents the complete evacuation of the galaxy, contrary 
to model MD-00-HI.  
 
In Fig. 7 the evolution of $M_{\rm ej,centr}$ and $M_{\rm 
ej,gal}$ for the two models MD-00-HI is compared .  As it is 
immediately apparent from this figure, much more ejecta is trapped 
into the galaxy in the model MD-00-HIbis during the first 200 Myr. 
This is due to the fact that the superbubble is expanding in a medium 
which is on average $\sim 3$ time denser 
and suffers larger radiative losses at the 
interfaces between hot and cold gas.  The ejecta is successively 
stripped together with the ISM.  After 230 Myr the metal content 
becomes similar to that of model MD-00-LO indicating that the galactic 
metal enrichment is not affected by the ram pressure in the long run. 
 
As shown in Paper I, models MD-45-HI and MD-90-HI are completely  
deprived of ISM after a time $\sim200-400$ Myr, comparable to the  
dynamical timescale of a galaxy group. Therefore
we did not make any galactic wind simulation in these models.

\section{Other models}

\subsection{Models SM}

Although during the ram pressure phase models SM-LO retain almost all  
of the ISM in the central region and approximately 40-60\% of the gas  
in the galactic region, the ISM is easily blown away by the galactic  
wind for any inclination angle (see Fig. 5).   
Winds in the models SM-HI have not been simulated because  
they are quickly deprived of gas by the ram pressure (cf. Paper I).  
 
\subsection{Models LG}  
\subsubsection{LG-LO} 
 
As for models MD-LO, also for the more massive galaxy models LG-LO
the evolution of the gas content is rather insensitive to the 
inclination angle $\theta$. Contrary to model MD-LO, instead, the onset of 
the SNe activity does not produce any significant decrease of $M_{\rm 
centr}$.  This is due to the fact that the central region for the LG 
model includes a larger volume, while the central hole carved by the 
wind is narrower ($\sim 800$ pc in radius) because of the denser 
ISM. 
 
The amount of metals trapped in the LG-LO models is generally larger 
than in models MD-LO. This is explained by the
more massive extraplanar halo of stripped ISM present 
in models LG-LO which efficiently contrasts the expansion 
of the superbubble. This effect is particularly dramatic
for $\theta=0$, when the trapped ejecta is maximum.
The presence of the halo 
increases the value of the critical luminosity needed by the wind to 
break out (e.g. Mac Low, 
McCray \& Norman 1989, Koo \& McKee 1992; see also Section 5.3). 
In the LG-LO-00 model the 
wind luminosity is quite close to the critical value, as discuss in 
section 5.3. As a consequence, part of the wind gas remains trapped 
in the halo and falls back on the galactic disk together with 
collapsing halo itself. 

A further reason for the larger values of $M_{\rm ej,centr}$ and
$M_{\rm ej,gal}$ in models LG-LO is the large size of the ISM disk,
which intercepts a larger amount of the metal rich gas pushed back
by the ram pressure. 
This effect is especially important for the LG-LO-90 model.

\subsubsection{LG-HI} 
 
As pointed out in section 2.3, the stripping simulations of our models 
in Paper I are adiabatic because, as stated in that paper, radiative 
losses are usually negligible. However, in the case of the model 
LG-00-HI this is not completely true.  In this model the ISM flare
influences more strongly the stripping evolution.  The 
extraplanar gaseous halo formed around the LG-00-HI model is dense enough
to make the radiative losses not negligible.
In order to make a self consistent model we run again LG-00-HI 
taking into 
account the radiative cooling {\it ab initio}, i.e. also during the 
ram pressure stage (model LG-00-HIbis). 
Given the smaller time step (due to the radiative time scale), we run 
this simulation only up to 500 Myr. 
In this case 
part of the halo is able to cool and to fall on the central
region of the galactic plane,
increasing $M_{\rm centr}$.  
The radiative halo is more clumpy 
and filamentary than the adiabatic halos forming in the previous 
models.
The massive, radiative extraplanar halo also explains why 
model LG-00-HIbis retains
much more metals than any other model. The cooled off and fallen
halo material is polluted with metals carried at high $z$ by the wind
and mixed with the halo gas.
This process is ultimately a consequence of the ram pressure and
may result in a different chemical evolution for galaxies
moving through an IGM with respect to isolated objects.

\begin{figure}  
\begin{center}  
\psfig{figure=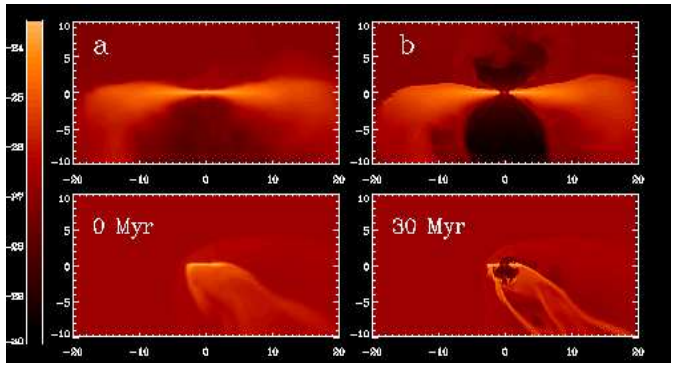}  
\end{center}  
\caption{Density distribution of the ISM for the two large $45^{\circ}$  
models with low ram pressure (LG-45-LO; upper panels) and high ram 
pressure (LG-45-HI; lower panel). The first and the second column 
refer to the initial condition at the beginning of the starburst (0 
Myr) and the end of it (30 Myr), respectively. The distances are given 
in kpc.} 
\end{figure} 

>From Figure 5 we see that the rate of mass loss for models
LG-45-HI and LG-90-HI is not significantly altered by the occurrence
of the starburst. Evidently the time evolution of $M_{\rm centr}$ 
and $M_{\rm gal}$ is regulated by the ram pressure alone.
This also 
indicates that in these models the gaseous halo, less massive
than in model LG-00-HI, does not radiate 
efficiently. In fact, because of the 
different geometry, the influence of the ISM flare is much less important 
for the LG-45-HI model and negligible for LG-90-HI model, where
no substantial gaseous halo is present.
In model LG-45-HI (shown in Figure 6), the halo surrounds only the
downstream side of the galaxy. 
 
The dependence on $\theta$ of the extraplanar halo formation induces a
dependence on $\theta$ of the evolution of $M_{\rm ej,centr}$ and 
$M_{\rm ej,gal}$ (see Fig. 8), because larger haloes lead to larger 
amounts of trapped ejecta. 
Moreover, a secondary reason for this
$\theta$-dependence is given by the fact that inclined galaxies have 
smaller radii (see Tab. 2) because of the ram 
pressure truncation, and thus are less able to retain metals
pushed back on the galactic disk.
 
The above arguments also explain why an analogous dependence on
$\theta$ is less pronounced in the case of lower ram pressure. In this
latter case the size of the gaseous galactic disk is less affected by
the inclination angle, and, more important, extends to a larger
$R$. As a consequence, the gas stripped from the galactic edge is less
dense and moves at higher $z$ over the galactic centre, opposing a
weaker contrast to the superwind expansion.  This is clearly shown in
Fig. 6 where the models LG-45-LO and LG-45-HI are compared. The effect
of the larger ram pressure is dramatic: the galactic size is extremely
reduced, and the lower superbubble lobe can not break out of the
gaseous halo, while the upper lobe is extremely distorted and bended
downstream.  

\begin{figure*} 
\begin{center}
\psfig{figure=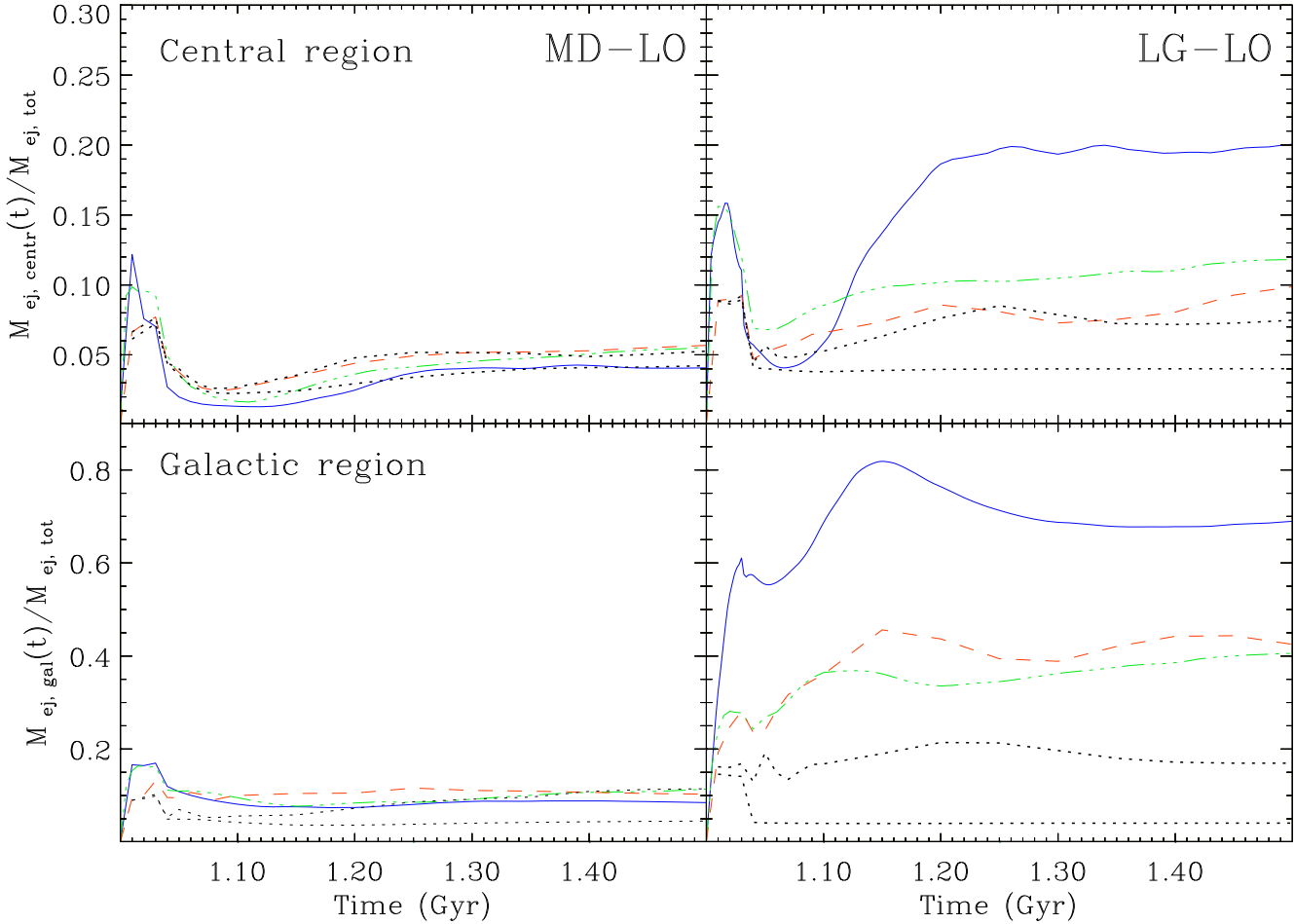} 
\end{center} 
\caption{Evolution of the tracer in the central region (upper panels)
and in the galactic region (lower panels). From left to right the
couples of lower and upper panels refer to MD and LG models,
respectively, in the case of lower ram pressure (LO).  Solid blue
lines: edge-on models; dashed-dotted green lines: $45^{\circ}$ models;
dashed red lines: face-on models. Higher and lower dotted lines in
each panel represent the FIELD and REST models, respectively.}
\end{figure*}

\begin{figure*}   
\begin{center}   
\psfig{figure=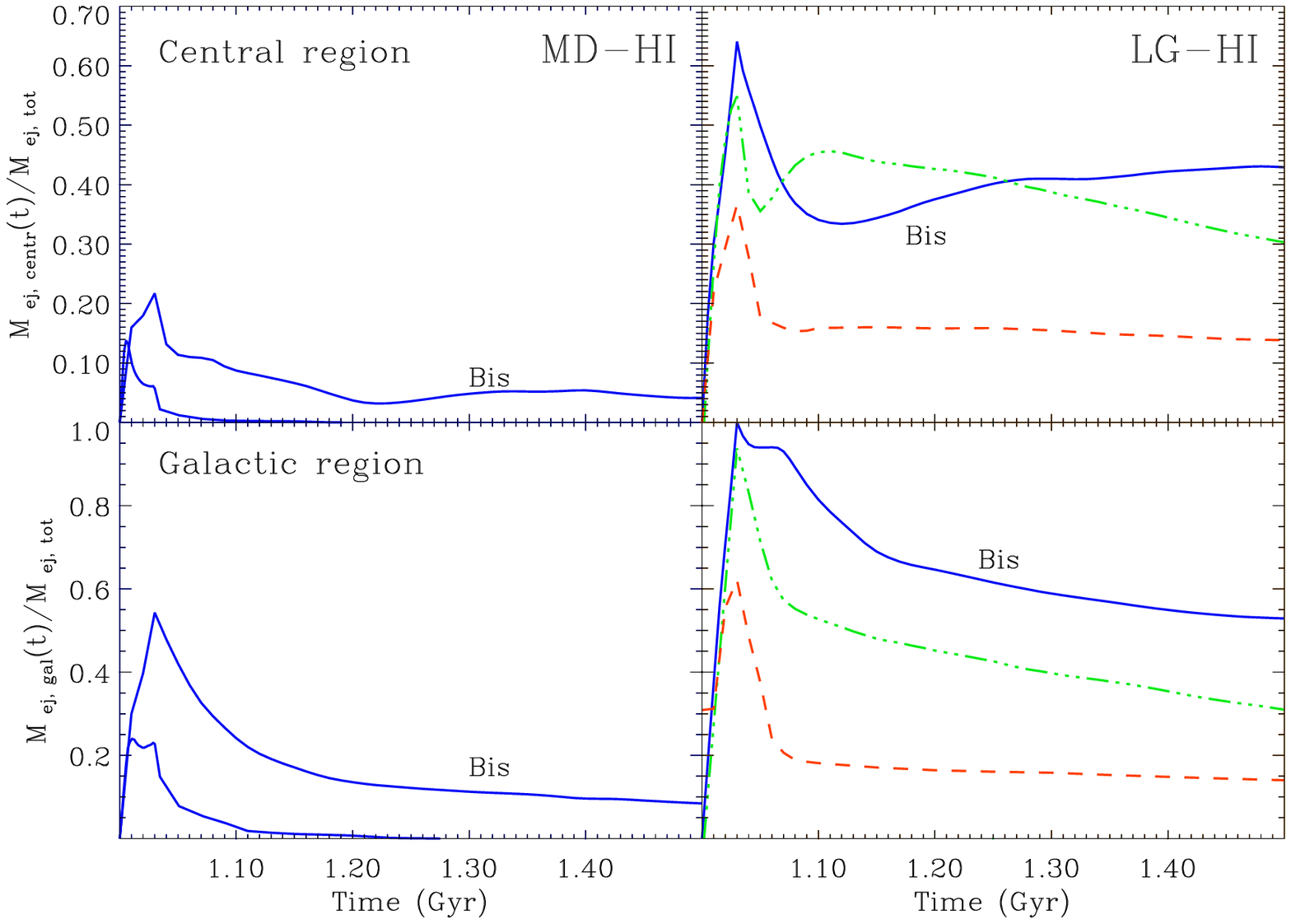}   
\end{center}  
\caption{Evolution of the tracer in the central region  
(upper panels) and in the galactic region (lower panels). From 
left to right the couples of lower and upper panels refer to MD 
and LG models, respectively, in the case of higher ram pressure (HI). 
Solid lines: edge-on blue models; dashed-dotted green lines: $45^{\circ}$ 
models; dashed red lines: face-on models.} 
\end{figure*}

\section{Discussion and conclusions} 
 
Here we briefly summarise and discuss the behaviour of the ISM and of 
the SN ejecta in our models.  For a better understanding of our 
results we also simulated the evolution of galactic winds occurring in the 
usual model galaxies but now assumed at rest 
relative to the IGM ($\rho_{\rm IGM}=2\times 10^{-28}$ g cm$^{-3}$,
$T_{\rm IGM} = 10^6$ K; REST models), or not surrounded by any IGM (FIELD 
models). The aim of these models is to obtain a more direct insight of 
the role played by the ram pressure comparing interesting quantities 
such as the ISM and ejecta mass content (see Figures 5 and 7)
in otherwise identical galaxies. The efficiency
of metal ejection is known to be sensitive to the details
of numerical simulations (cf. D'Ercole \& Brighenti 1999
and MacLow \& Ferrara 1999), 
and a consistent comparison must be done among
similar models.

\subsection{ISM evolution}

The ISM of SM galaxies subject to the high ram pressure 
is quickly dragged away and therefore we did not 
simulate winds for SM-HI models.
The low ram pressure strips only $30-50$ \% of the original ISM of small
galaxies (SM-LO models). However, the
starburst powered wind completely removes all the gas in $\approx 100$ Myr, 
a result obtained also by the galaxy at rest. Evidently the ram pressure
has a negligible influence on the galactic wind evolution for these
models. 
For more massive 
galaxies the occurrence of a starburst does not 
influence significantly 
the ongoing mass loss due to the ram pressure in the MD-LO 
models and can only anticipate the complete removal of gas
(e.g. for model MD-00-HI). 

For the more massive LG galaxies the situation is complicated by 
the radiative nature of the gaseous halo which develops around the 
galactic disk, especially in the edge-on model with the larger ram 
pressure (LG-00-HI). 
Cooling of the ablated gas makes $M_{\rm centr}$ to increase 
by a factor of few; $M_{\rm gal}$, instead, is not affected. 
As for the other models, the
occurrence of a starburst in models LG does not influence the
time evolution of $M_{\rm gal}$, which is determined by the ram 
pressure in both the HI and LO cases. We stress that in these models, 
as well as in any other model in which the galaxy is not rapidly 
deprived of gas by the wind, the ISM loses memory of the
starburst after a few tens of Myr and recovers a distribution
similar to the initial one. At this point the galaxy is in principle
ready for another possible starburst episode.

\begin{figure*} 
\begin{center} 
\psfig{figure=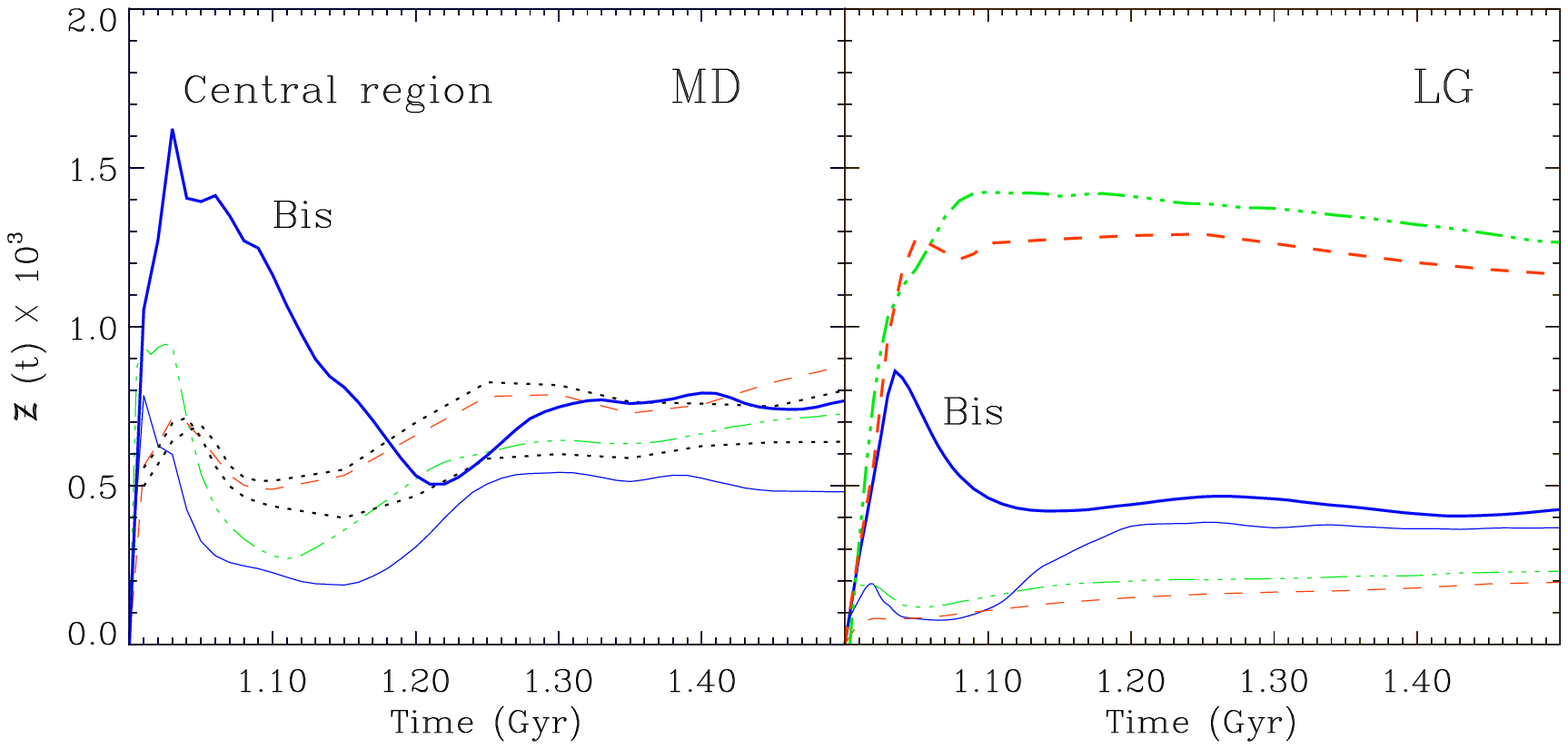}
\end{center} 
\caption{Evolution of ${\cal Z}$ (see text) in the central region.
The left and right panel refer to MD and LG models, respectively.
Solid blue lines: edge-on models; dashed-dotted green lines:
$45^{\circ}$ models; dashed red lines: face-on models.  Higher and
lower dotted lines in the right panel represent the FIELD and REST
models, respectively.  Light and heavy lines refer to LO and HI ram
pressure models, respectively.}  
\end{figure*}

In conclusion, our models show that
the galactic wind either disrupts the whole ISM
or has a negligible effect on the ISM content,
which in turn is regulated mainly by the ram pressure.
When $\rho_{\rm IGM} v^2_{\rm IGM}
\sim P_0$, where $P_0$ is the central ISM 
thermal pressure, the ram pressure stripping and the wind act
together to increase the ISM 
removal rate.
Many parameters regulate the ISM dynamics in this case
(intensity and duration of the ram
pressure, wind mechanical luminosity, potential well depth, 
initial amount of the
ISM, value of the inclination angle $\theta$) and numerical
simulations are needed to understand the ISM behaviour of a specific
model. 

\subsection{Ejecta evolution}

For our low mass galaxies (models SM) the wind expels
the whole ISM and no ejecta remains trapped into them.
The more
massive models with the lower ram pressure (MD-LO models) retain 
5\% of the total ejecta mass in the central region, and 10\% in the 
whole galaxy. These are essentially the same values obtained by the 
analogous REST model, while for the analogous FIELD model the trapped 
quantities are lower by a factor of $\sim 2$.
We thus conclude that for these models the ram pressure
has little effect in the entrapment of the SN metals, while
the presence of a relatively high pressure ambient gas
helps this task. These results are in qualitative agreement
with those in Murakami \& Babul (1999).
A quantitative comparison
with the models by Murakami \& Babul however is not possible.
In fact, having these authors considered a spherical galaxy,
the (metal rich) superbubble preserves a spherical shape 
when the ram pressure is negligible, and collapses as
a whole. Thus, all the metals produced by the starburst
are retained by the galaxy.
On the other hand, in our flattened galaxies a large fraction
of the ejecta is always lost through the superbubble breakout.

With the larger ram pressure, the ejecta and the ISM in the edge-on 
model MD-00-HI (the only model which keeps some ISM at the end of the
ram pressure phase) are stripped after $t\sim 200$ Myr
(Figure 8). For $t < 100$ Myr, however, the time evolution of
the ejecta masses in both the central and galactic region is
remarkably similar to that of model MD-00-LO.  
If the starburst occurs
when the values of $M_{\rm centr}$ is high (model MDbis) more 
ejecta gets trapped, peaking at $t \simeq 30$ Myr with values of 20\% 
and 55\% in the central and galactic region, respectively. Then it 
decreases to 5\% and 10\% in the two regions. 
 
The deeper potential well and the larger ISM amount of the LG-LO models
are not the direct cause of the larger
retention of metals with respect to the MD-LO galaxies, 
as can be verified comparing 
the FIELD and REST models for the LG and MD galaxies.
Thus, the larger fraction of trapped metals 
occurring in the moving LG models is due to the presence of the 
extraplanar gaseous halo which develops in these models (in the 
edge-on case in particular), especially with the larger ram 
pressure. For the weaker ram pressure we find that the edge-on model 
keeps 20\% of the ejecta in the central region at the end, though goes 
through a minima at $t$=70 Myr with 5\%. The rise of $M_{\rm 
ej,centr}$ after this time is due to the collapsing halo gas polluted 
by the metals carried by the wind. The halo is also responsible of the 
rather high value ($\sim 70$ \%) of the mass of the ejecta trapped in 
the whole galactic region of the LG-00-LO model. For different values 
of $\theta$, the halo is less developed, and less metals are trapped: 
around 10\% in the central region, and $\sim 40$ \% in the galactic 
region, with little differences for various values of $\theta$. 
 
For LG-HI models the halo is even more influential. Now the ejecta
trapped in the central region is substantial: $\sim 43$\% for
$\theta=0^{\circ}$,  $\sim 30$\% for $\theta=45^{\circ}$, and $\sim 14$\% for
$\theta=90^{\circ}$. Very little ejecta is exterior to the central
region, so the values of $M_{\rm ej,gal}$ are very similar to those of
$M_{\rm ej,centr}$.
Despite the presence of the more massive gaseous halo,
$M_{\rm ej,gal}$ in the LG-HI models becomes somewhat
lower than in the LG-LO
models because of the stronger ram pressure which continuously erodes
the polluted ISM.

We conclude that, contrary to the ISM dynamics, 
the amount of the SN ejecta trapped into 
the galaxy results to be more affected by the action of the ram 
pressure. Part of the ejecta expelled by the superwind is pushed back 
onto the galaxy by the incoming IGM or remains trapped in the 
surrounding halo, and the fraction of metals retained by a moving 
galaxy can be up to 3 times (depending on the value of $\theta$) 
larger than that retained by a galaxy at rest. This trend
is opposite to that found by Murakami \& Babul (1999) in their
suite of models, where they also investigated the effect of
group-like ram pressure on galactic 
winds. This discrepancy is due to the assumed
spherical ISM distribution of their models which allows the complete
retention of the metals in the simulations without ram pressure.

Overall, the main conclusion by De Young \& Heckman (1994),
D'Ercole \& Brighenti (1999) and Mac Low \& Ferrara (1999)
are not changed by the ram pressure stripping:
for galaxies of size comparable to SM and MD
only a
very small fraction on the metals (of the order of $\approx 10$ \%) remains
trapped in the ISM.
Only the large
models LG can trap much more ejecta, an effect of the relatively large 
amount of gas located at high $z$.
This is conveyed there by the ram pressure and depends on
the flare in the ISM distribution (we discuss this point in section
5.4).
The (substantial) metal mass not incorporated in the ISM at $t=500$ Myr
is definitively lost by
the galaxies and enriches the IGM, on spatial scales on the order of
$30-50$ kpc.
The general trend shown by our models is that the ram pressure
increases
$M_{\rm ej,centr}$ and $M_{\rm ej,gal}$.

To characterise the 
pollution degree we calculate the average ejecta fraction in the 
central region defined as ${\cal Z}=M_{\rm ej,centr}^{\rm cold}/\ 
M_{\rm centr}$, where we consider only the cold fraction ($T<10^5$ K) 
of the ejecta which is effectively trapped in the ISM. 
In order to obtain the abundance of a specific X-element, one has to 
scale the ${\cal Z}$ value by a factor $Z_{\rm X,ej}$ representing the 
abundance of the X-element in the SN ejecta. Focussing on iron and 
oxigen, we assume $Z_{\rm Fe,ej}=4.4 
\times 10^{-3}$ and $Z_{\rm O,ej}=4.4 \times 10^{-2}$, and obtain 
$Z_{\rm Fe}/Z_{\rm Fe,\odot}=3.4 {\cal Z}$ and $Z_{\rm 
O}/Z_{\rm O,\odot}=4.6 {\cal Z}$, respectively (cf. D'Ercole \& 
Brighenti 1999 for more details). 

In Fig. 9 we show the evolution of $\cal Z$ in our models, and we
note that in general larger ram pressures lead to an higher metal 
enrichment. 
Thus, the ability to retain metals appears to be sensitive 
to the parameters regulating the interaction ISM-IGM,
and this may explain part of the observed scatter in the 
metallicity-luminosity relation (e.g. Lee, McCall \& Richer 2003).
We may also expect that dwarf irregulars in relatively high ram pressure
environments have systematically larger metallicity than the
field counterpart. Marginal evidence for such a trend has been
claimed by Vilchez (1995) in his study of Virgo Irregulars.
Elevated oxygen abundances for Virgo dwarfs has ben suggested also by
Vilchez \& Iglesias-Paramo (2003).

\subsection{The gaseous halo} 
 
As discussed above the gas which accumulates at large $z$
is a result of the ram pressure of the IGM with a flared ISM distribution.
Its presence allows us to compare our results to those by Silich \& 
Tenorio-Tagle (2001) and Legrand et al. (2001). These authors make a 
systematic study on the critical superwind luminosity needed for the
superbubble to break out. They conclude that star-forming dwarf galaxies 
must have an extended gaseous halo in order retain their metals and 
enhance their abundances. A direct comparison between our edge-on 
models and the models by Silich \& Tenorio-Tagle is prevented by the 
fact the galactic models are build following different criteria; the 
shape of the gaseous halo is also different. However, we consider two 
models by Silich \& Tenorio-Tagle, their models M800.100 and M900.100, 
which have values of $M_{\rm g}$ and $M_{\rm h}$ similar to our MD and LG 
models, respectively. Moreover, the column density of the gaseous halo 
of these models during the edge-on stripping is also similar to the 
column density of the halo of the two quoted models by Silich \& 
Tenorio-Tagle. Following these latter authors, the wind luminosity of 
our models is much larger than the critical luminosity needed by the 
superbubble to break out in the MD models. Thus 
the wind easily moves far from the galaxy bringing away most of the
SNII ejecta. On the contrary, the wind luminosity would be only
marginally sufficient to allow the breakout of the superbubble in the
LG model. Actually, our simulations show that the superbubble breaks
only marginally in this case, and the larger fraction of ejecta
trapped in the galaxy (Fig. 7-8) indicates a more substantial role of
the extraplanar gaseous halo.

\subsection{Limitations of the models and future work} 
 
In evaluating the above results one must bear in mind the two caveats
discussed above: $i$) the numerical diffusion, and $ii$) 
the flared ISM distribution. 
The numerical diffusion prevents the possibility to give 
an accurate {\it quantitative} estimate of the ejecta 
mixed with the ISM. 
However, we believe that the {\it relative} differences 
among the models are still meaningful, and thus ram pressure may indeed
increase the metal enrichment of the ISM due to a starburst. 

The metal enrichment in our models is also influenced by the ISM flare.
This influence is particularly evident for
large galaxies moving edge-on through the IGM (LG-00 models), where 
a gaseous halo form around the galaxy, affecting the superwind 
expansion. The flare in our models derives from the assumption of 
isothermal ISM, which requires a rotation curve independent of 
$z$ in order to assure hydrostatic equilibrium (e.g. Tassoul 1978).
 
The real presence and the effective extension of flares in galaxies is 
still an open question, although evidence for flares
in galaxies is claimed by several authors (e.g. Brinks \& Burton 1984, 
Burton 1988, Olling 1996, Matthews \& Wood 2003). Simulations 
analogous to those presented here but without flare in the initial ISM 
distribution would be interesting. 
We are currently devising non-isothermal models for the ISM 
to evaluate
the real influence of the ISM distribution on the interaction with the 
IGM distribution. 
 
\section*{Acknowledgements}
We are grateful to the referee, Arif Babul, for a number of
suggestions which improved the presentation of the paper.
Many thanks to Renzo Sancisi for useful discussions on the ISM 
distribution. We acknowledge financial support from National 
Institute for Astrophysics (INAF). The simulations were run at the 
CINECA Supercomputing Centre with CPU time assigned thanks to 
INAF-CINECA grant.

{}

\label{lastpage}
\end{document}